\documentclass[natbib209]{emulateapj}
\newcommand{\hi}{\mbox{H{\small I}}}

\shorttitle{Dust in the Extremely Metal Poor Galaxy I Zw18}
\shortauthors{Herrera-Camus et al.}

\begin{document}

\title{Dust-to-Gas Ratio in the Extremely Metal Poor Galaxy I Zw 18}


\author{Rodrigo~Herrera-Camus\altaffilmark{1}, David~B.~Fisher\altaffilmark{1}, Alberto~D.~Bolatto\altaffilmark{1}, Adam~K.~Leroy\altaffilmark{2}, Fabian~Walter\altaffilmark{3}, Karl.~D.~Gordon\altaffilmark{4}, Julia~Roman-Duval\altaffilmark{4},Jessica~Donaldson\altaffilmark{1}, Marcio~Mel\'endez\altaffilmark{1}}
\author{John~M.~Cannon\altaffilmark{5}}

\altaffiltext{1}{Department of Astronomy, University of Maryland, College Park, MD 20742, USA.}
\altaffiltext{2}{National Radio Astronomy Observatory, 520 Edgemont Road, Charlottesville, VA 22903-2475, USA.}
\altaffiltext{3}{Max-Planck-Institut f\"{u}r Astronomie, Konigst\"{u}hl 17,  D-69117 Heidelberg, Germany.}
\altaffiltext{4}{Space Telescope Science Institute, 3700 San Martin Drive, Baltimore, MD 21218.}
\altaffiltext{5}{Department of Physics \& Astronomy, Macalester College, 1600 Grand Avenue, Saint Paul, MN 55105, USA.}

\begin{abstract}
The blue compact dwarf galaxy I~Zw~18 is one of the most metal poor
systems known in the local Universe (12 + log(O/H) $=$ 7.17). In this work we
study I~Zw~18 using data from {\it Spitzer}, {\it Herschel Space Telescope} and 
IRAM Plateau de Bure Interferometer.
Our data set includes the most sensitive maps of I~Zw~18, to date, 
in both, the far infrared and the CO $J=1\rightarrow0$ transition.
We use dust emission models to 
derive a dust mass upper limit of only M$_{dust}\leq1.1\times10^4$ M$_{\odot}$ 
($3\sigma$ limit). This upper limit is driven by the non-detection at 160~$\mu$m,
and it is a factor of $4-10$ times smaller than previous estimates
(depending upon the model used).
We also estimate an upper limit to the total dust-to-gas mass ratio of
M$_{Dust}$/M$_{gas}\leq5.0\times10^{-5}$.
If a linear correlation between the dust-to-gas mass ratio and metallicity (measure as O/H) were to hold,
we would expect a ratio of 3.9$\times10^{-4}$. 
We also show that the infrared SED is similar to that of starbursting systems.
\end{abstract}


\section{Introduction}

The link between dust-to-gas mass ratio (DGR) and heavy element
abundance (metallicity) in galaxies remains an open issue
\citep[e.g,][]{rhc_lisenfeld98,rhc_edmunds01,rhc_hunt05}.
Specifically, in very low metallicity systems ($12+\log({\rm
O/H})\lesssim8$) it is unclear how the DGR scales with
metallicity. Models considering dust destruction by supernovae
\citep{rhc_hirashita02} or mass outflows from the galaxy
\citep{rhc_lisenfeld98} predict a nonlinear relation. On the other
hand, if the fraction of metals incorporated in the dust is constant
\citep{rhc_james02}, we expect a linear
relation between DGR and metallicity, in a sense that the ratio decreases 
as metallicity decreases.  Measurements of DGRs over a range of 
metallicity are necessary to better constrain this relationship.

The blue compact dwarf galaxy I~Zw~18 has one of the lowest nebular
metallicities measured to date. \cite{rhc_skillman93} measure an
oxygen abundance of 12+log(O/H)~=~7.17. This is 3.2\% of the solar 
abundance (using the scale of \citealt{rhc_asplund09}). 
Most local universe galaxies have 12+log(O/H)~$\sim8.5$
\citep[e.g.,][for SINGS]{rhc_moustakas10}, and the 
Milky Way has 12+log(O/H)$\sim8.7$ \citep{rhc_baumgartner06}. 
I~Zw~18 therefore represents the extreme low end of the 
metallicity range in the local universe and is thus
a key datum for understanding the relationship between DGR and
metallicity. 

The dust mass of I~Zw~18 is poorly known. 
Typical galaxies of similar morphology (blue compact dwarfs)
have dust masses that range between 10$^{3}-10^{5}$~M$_{\odot}$, 
with DGR ranging between $10^{-3}-10^{-5}$ \citep{rhc_lisenfeld98}.
Using H$\alpha$/H$\beta$ flux
ratios as a dust tracer, \citet{rhc_cannon02} find a total dust mass
for I~Zw~18 of $\sim7-10~\times~10^{3}$~M$_{\odot}$\footnote{We scale
\citet{rhc_cannon02} and \citet{rhc_engelbracht08} result by 
a factor of (18.2/12.6)$^{2}$ and
\cite{rhc_galametz11} result by a factor of (18.2/13)$^{2}$ to account
for the differences in assumed distances. We note that our final
result, the DGR, is distance independent.} by assuming a linear
scaling between DGR and metallicity (as measured by O/H). \cite{rhc_engelbracht08},
using {\em Spitzer} data limited by a non-detection at 160~$\mu$m,
measure an upper limit for the dust mass of  $4.2\times10^4$~M$_{\odot}$.
A more recent study of a large sample by \cite{rhc_galametz11} 
uses previously published {\em Spitzer} and SCUBA data to constrain
the dust mass of I~Zw~18 to be $\lesssim1.1\times10^{5}~$M$_{\odot}$
and the DGR to $\lesssim4.5\times10^{-4}$.

I~Zw~18 contains intense radiation fields stemming from
active star formation. It therefore provides a nearby
testing ground to probe the physics of distant primeval
sources. Previous studies using {\it Spitzer}
\citep{rhc_engelbracht08,rhc_wu07} show that its continuum emission
from 15 to 70~$\mu$m has a slope characteristic of a starburst galaxy
of solar abundance. Moreover, its mid-infrared spectrum from 5 to
36~$\mu$m shows no detectable emission from polycyclic aromatic
hydrocarbons (PAHs). Such low abundance of PAHs is likely the
consequence of a high radiation field in combination with the low
metallicity of the source.

In this paper we estimate the DGR for I~Zw~18. We use previously unpublished 
{\it Spitzer Space Telescope} and archival {\it Herschel Space
Observatory}\footnote{Herschel is an ESA space observatory with
science instruments provided by European-led Principal Investigator
consortia and with important participation from NASA.} continuum
observations, combined with dust emission models
and a gas mass \citep{rhc_vanzee98} to constrain the radiation field
intensity, temperature, dust mass and DGR in I~Zw~18. Throughout this
paper we assume a distance of 18.2 Mpc \citep{rhc_aloisi07}. 
Revisions to this distance will affect our dust and gas masses, but not the DGR.

\begin{figure*}
\epsscale{1.0}
\plotone{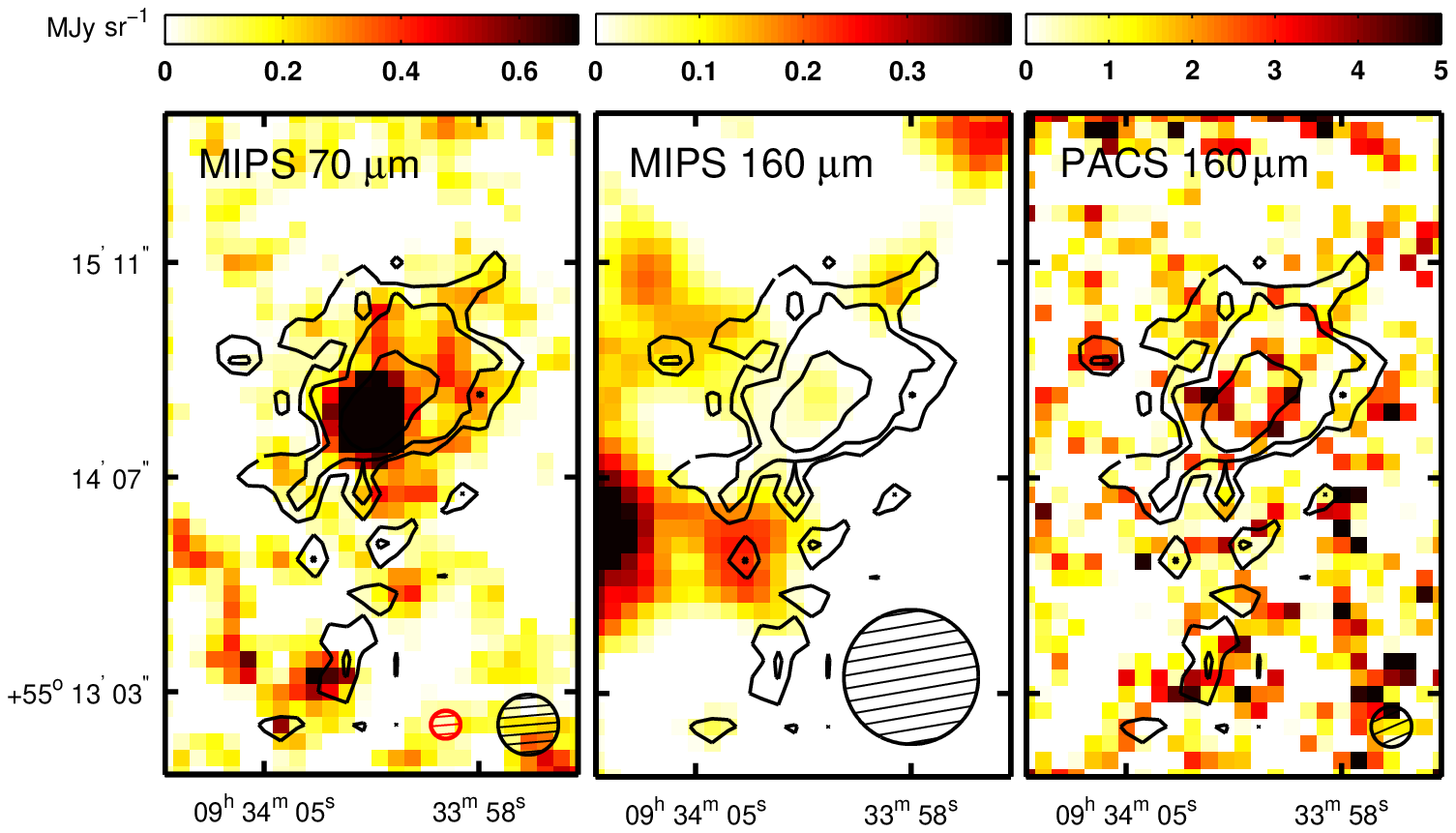}
\caption{ The left two panels show the {\it Spitzer} images of I~Zw~18 at 70
and 160~$\mu$m. The right panel shows the {\it Herschel} PACS image at
160~$\mu$m. Overlaid as black contours is the \hi\ column density
distribution from \cite{rhc_vanzee98} using the VLA.  The contours are
0.7, 1.4 and 5~$\times$~ 10$^{20}$~cm$^{-2}$. The black circle in the
bottom right corner of each panel corresponds to the respective beam size
of the FIR observations. The smaller circle in the first panel
corresponds to the beam size of the \hi\ observations.
At 70~$\mu$m, the bulk of the emission coincides with the \hi\ contours
and the diffuse emission extends preferentially towards the
NW. I~Zw~18 is undetected in both 160~$\mu$m maps. 
\label{map_panel}}
\end{figure*}

\begin{figure*}
\epsscale{1}
\plotone{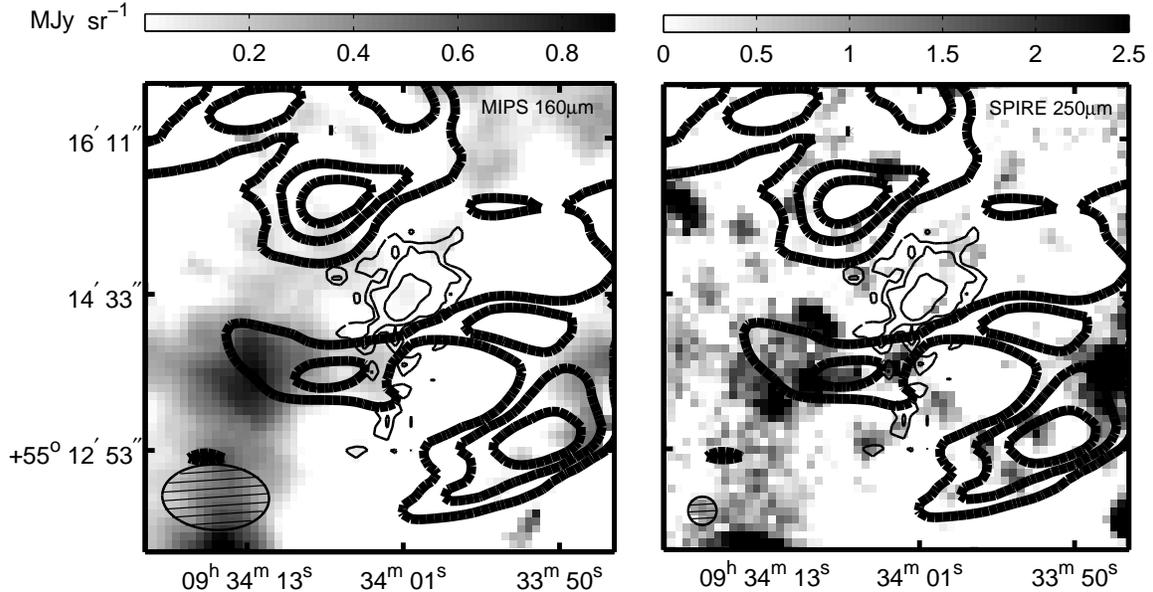}
\caption{
{\it Spitzer} 160~$\mu$m ({\it left}) and {\it Herschel} 250~$\mu$m ({\it right}) 
maps of a $\sim$5~$\times$~5~arcmin field
around I~Zw~18. The thick black contours show the
VLA observations of the Galactic \hi\ foreground emission at 
2, 4 and 6$\sigma$ significance level.
The thin black contours are the same as shown in Fig.~\ref{map_panel}.
The VLA beam size is $\theta=67.3\arcsec\times41.3\arcsec$
and is shown in the corner of the left panel. The SPIRE 250~$\mu$m beam is shown
in the right panel. The bulk of the 160~$\mu$m emission,
located south-east of our object, fragments into at least three point sources
in the 250~$\mu$m map, which has better spatial resolution. 
There is no correlation between the far infrared and 
the Galactic \hi\ foreground emission, 
suggesting that the confusion is dominated by background
galaxies. \label{confusion}}
\end{figure*}

\section{Methods}

\subsection{Observations}

We use a variety of data from several different facilities to map the far infrared, 
sub millimeter, millimeter and radio wave emission of I~Zw~18. 
Therefore our data set uses the following observatories and instruments: 
 {\it Spitzer} Multiband Imaging Spectrometer (MIPS,  \citealt{rhc_rieke04}); {\it Herschel} 
 Photodetector Array Camera and Spectrometer (PACS, \citealt{rhc_poglitisch10}); 
 {\it Herschel} Spectral and Photometric Imaging Receiver (SPIRE, \citealt{rhc_griffin10}); 
IRAM Plateau de Bure Interferometer (PdBI) and Very Large Array (VLA). In this section
we will briefly describe new observations.

{\textit{\textbf{Spitzer:}}} We observed I~Zw~18 at 70 and 160~$\mu$m using
MIPS in photometry mode as part of a
cycle 5 proposal (P.I. A. Bolatto, AOR: 22369536). 
The total observation time was 8
hours. The reduction of these images very closely follows the
procedure described in \citet{rhc_gordon07} and
\citet{rhc_stansberry07} for the 70 and 160~$\mu$m map respectively.

{\textit{\textbf{Herschel}}}{\bf /PACS:} We use archival 70 and
160~$\mu$m observations from {\it Herschel} . The observations were
taken with PACS using the Large Scan Map mode as part of the {\it
  Herschel} Guaranteed Time Key Program, Dwarf Galaxy Survey (P.I.
S. Madden, obs. ID: 1342187135/36). The scan maps were taken at
90$^{\circ }$ angles from one another at the medium scan speed
($20~\arcsec$~s$^{-1}$) and then combined together in order to reduce
the noise caused by streaking along the scan direction.  The scan leg
length is $4.0~\arcmin$ and the total on source time for the combined
images was 192~s.

Unlike {\it Spitzer} data, the methods to reduce PACS data are still
evolving significantly. Therefore, we reduce the data in two separate
ways.  We first use Herschel Interactive Processing Environment (HIPE)
v4.2 with the standard pipeline scripts. We also process the data up
to level 1 in HIPE v7.  We use the standard pipeline, which includes
pixel flagging, flux density conversion, and sky coordinate
association for each pixel of the detector.  At this stage, the PACS
timelines are still affected by $1/f$ noise and baseline
drifts.  In order to subtract the baseline, remove glitches, and
project the timelines on the final map, we applied the scanamorphos
algorithm \citep{rhc_roussel12} to the level-1 PACS timelines.

{\textit{\textbf{Herschel}}}{\bf /SPIRE:} We use archival {\it
  Herschel} Spectral and Photometric Imaging Receiver (SPIRE;
\citealt{rhc_griffin10}) photometric observations at 250, 350, and 500
$\micron$. Observations were made in the large map mode with the
nominal scan speed of 30~arcsec~s$^{-1}$ and the cross-scanning method
as part of {\it Herschel} Science Demonstration Phase (PI: S. Madden;
obs. ID 1342188663).  Data reprocessing was carried out in HIPE using
the standard large map pipeline with the latest SPIRE calibration tree
available\footnote{We used HIPE 8.1 and the SPIRE calibration tree
  v. 8.1.} which includes deglitching the timeline data, flux
calibration and various corrections. After removal of a linear
baseline, images were made using the standard naive mapper. The final
maps are in units Jy beam$^{-1}$ with pixel scales of 6, 10 and
14~arcsec at 250, 350 and 500~$\mu$m, respectively as described in the 
SPIRE Data Reduction Guide.

{\bf PdBI:} We present new observations of the CO $J=1\rightarrow0$ 
transition in I~Zw~18 using the IRAM Plateau de Bure Interferometer 
as project t027 (P.I. A. Leroy). The data were observed on 24, 27, and 28 
September 2009 using the "5Dq" configuration, meaning that 5 telescopes 
were operational and that the array was in a compact configuration. 
They data is calibrated in the standard way in December 2009 
using the PdBI pipeline implemented in the CLIC and MAPPING packages 
of GILDAS. The effective time on source was 12.5 hours after flagging during 
the pipeline run. The effective bandwidth was $\sim 850$~MHz, or about 
2200~km~s$^{-1}$ with native resolution $\sim2.5$~MHz (6.5~km~s$^{-1}$).
We do not detect CO emission. At 26~km~s$^{-1}$ velocity resolution
we achieved an RMS noise of 1.26~mJy~beam$^{-1}$, implying a $4\sigma$ 
flux upper limit for a point source of 0.131~Jy~km~s$^{-1}$.

{\bf VLA:} 
The observations used to construct the \hi\ map
are described in \cite{rhc_vanzee98}. We obtained two hours of Rapid
Response 21~cm VLA observations (project 08B-246; P.I. A.
Bolatto)  to evaluate the Galactic foreground contribution. This contribution can be estimated
by measuring the HI column density toward I~Zw~18 and convert it to dust emission
using typical high-latitude Galactic ratios \citep[e.g,][]{rhc_boulanger96}. 
The observations were obtained
during the move between CnD and D configuration with a synthesized
beam size of $67\arcsec\times41\arcsec$, and a native resolution of 6.1~kHz (1.3~km~s$^{-1}$).
Data was reduced in AIPS using the standard procedure and calibrations, and care was taken to
remove the baselines affected by the frequency aliasing problems due to the VLA-JVLA transition.
At 10.3~km~s$^{-1}$ velocity resolution we achieved an RMS noise of 1.1~mJy~beam$^{-1}$, implying 
an \hi\ column density of N$_{\rm HI} = 2.4\times10^{18}$~cm$^{-2}$.
Galactic neutral hydrogen emission was observed in the central 30~km~s$^{-1}$ of the passband.
Nonetheless, even after spatially filtering the 160~$\mu$m MIPS map to approximately match
the $uv$ coverage of the VLA the correlation between the high resolution HI column density
and the 160 $\mu$m surface brightness remained extremely low (Fig.~\ref{confusion}) , showing
that most of the emission present in the 160 and 250 $\mu$m images is not due to the high-latitude
Galactic foreground.

\subsection{Photometry} 

In Fig.~\ref{map_panel} we present the {\it Spitzer} MIPS 70 and
160~$\mu$m maps and the {\it Herschel} PACS 160~$\mu$m image for
I~Zw~18.  Overlaid on each image is the \hi\ column density
distribution observed by \cite{rhc_vanzee98}. The \hi\ contours
correspond to 0.7, 1.4 and 5~$\times$~10$^{20}$~cm$^{-2}$ enclosing
98, 96 and 78\% of the total flux at 70~$\mu$m.  We use the MIPS
70~$\mu$m map over the PACS 70~$\mu$m because it has a much better
surface brightness sensitivity ($0.17$ versus $2.87$ MJy~sr$^{-1}$),
yielding a better signal to noise.  The bulk of the 70~$\mu$m emission
coincides with the \hi\ column density maximum. This peak also
coincides with the location of active star forming regions observed by
\citet{rhc_cannon02} using the {\it Hubble Space Telescope} (HST). The
diffuse component at 70~$\mu$m extends preferentially $\sim$3~kpc
north-west from the peak. We subtract the background emission measured
in a region free of sources. We integrate the flux using a circular
aperture with radius of 45'' centered at the peak of the 70~$\mu$m
emission and applying an aperture correction factor of 1.13
(determined by integrating over the point spread function and
compatible with those in the MIPS Instrument Handbook).  The
calibration error on {\it Spitzer} is about 5\% at 70$\mu$m.  We
estimate a photometry error of 1.7~mJy by adding in quadrature the
calibration uncertainty and the background noise We measure a total
flux density of 33.6$\pm$1.7~mJy at 70~$\mu$m. Our flux value is
consistent with the 70~$\mu$m flux measured by
\cite{rhc_engelbracht08} of 34.9$\pm$4.79~mJy.

The {\it Spitzer} 160~$\mu$m map is confusion-limited. The bulk of the
emission is associated with sources outside the \hi\ emitting region of
I~Zw~18. Although the Galactic latitude of I~Zw~18 is $\sim$44$^{\circ}$, 
it could be possible that a significant source of confusion were Galactic cirrus.
We use the VLA \hi\ observations to explore this possibility.
Fig.~\ref{confusion} shows the MIPS 160~$\mu$m
and the SPIRE 250~$\mu$m map of a $\sim$5$\times$5~arcmin field 
around I~Zw~18. The thick lines represent the \hi\ foreground emission
from VLA. The thin lines represent the \hi\ emission of I~Zw~18.
It is clear from visual inspection that the maxima of
the \hi\ foreground and the 160~$\mu$m and 250~$\mu$m 
emission are not coincident. 
We find a Pearson correlation coefficient close to
zero. In most of the {\it Spitzer} confusion-limited images at
160~$\mu$m the confusion is mainly due to faint unresolved background
sources \citep{rhc_Dole04}. The bulk of the 160~$\mu$m emission,
located south-east of our object, fragments into at least three point sources
in the 250~$\mu$m map, which has finer spatial resolution. 
The 160~$\mu$m peak also coincides with
several background galaxies in deep {\it B} and {\it R}-band images (S.~
Janowiecki, private communication). The difference between the 
160 and 250~$\mu$m maps is consistent with what one would observe if the peak of the emission 
at a 160~$\mu$m is associated with background galaxies.

The background contamination and the absence of
correlation between the 160/250~$\mu$m emission and the \hi\ foreground emission
makes it impossible to recover the flux associated with I~Zw~18. Thus, we
use an annular sector around the galaxy to measure a one-sigma
surface brightness sensitivity of 0.18~mJy~sr$^{-1}$
that includes the effects of confusion. We estimate the flux upper
limit multiplying this value by the area associated with the
1.4$\times10^{20}$~atoms~cm$^{-2}$ \hi\ contour that encloses 96$\%$
of the 70~$\mu$m flux and 62$\%$ of the \hi\ mass. To find
the aperture correction factor associated to this area, we can approximate
the contour using a circular aperture of 48\arcsec\ radius.
This corresponds to an aperture correction factor of 1.6 at 160~$\mu$m.
After applying the aperture correction factor, we obtain a
corresponding 3$\sigma$ flux upper limit of 40.5~mJy. This new upper
limit is a factor of $\sim$2 lower than the previous upper limit
published by \cite{rhc_engelbracht08}.

The {\it Herschel} PACS image at 160~$\mu$m also fails to detect
I~Zw~18. In this case, however, the image is not
confusion-limited. The $1\sigma$ surface brightness sensitivity is
1.8~MJy~sr$^{-1}$. If we assume that the emission from I~Zw~18 is
compact on $12\arcsec$ scales, the corresponding 3$\sigma$ flux upper 
limit integrating over the $12\arcsec$ beam and applying an aperture 
correction factor of 1.32 is 27.2~mJy. 

We will work with the PACS 160~$\mu$m flux upper limit of 27.2~mJy
for the rest of the paper. However, the upper limit measured from the 
PACS data relies upon the assumption that the source is compact.
If the 160~$\mu$m emission is significantly 
extended over scales larger than $12\arcsec$ ($\sim$1~kpc),
it may be more appropriate to use the MIPS 160~$\mu$m upper limit
of 40.5~mJy.

Finally, the SPIRE maps at 250, 350 and 500~$\mu$m show no detection
of I~Zw~18. 
From these images we measure a
surface brightness sensitivity using an annular sector 
around the source. We then apply aperture corrections and point source color
corrections assuming $\beta=1.5$ ($\beta$ in $f_\nu \propto
\nu^\beta$) described in the SPIRE Photometry Cookbook 
(Bendo, G. J. and the SPIRE-ICC, 2011)
We measure 3$\sigma$ flux upper limits 
of 22.2, 23.9 and 25.4~mJy at 250, 350 and 500~$\mu$m, respectively.




\section{Results}

\begin{figure}
\epsscale{1.0}
\plotone{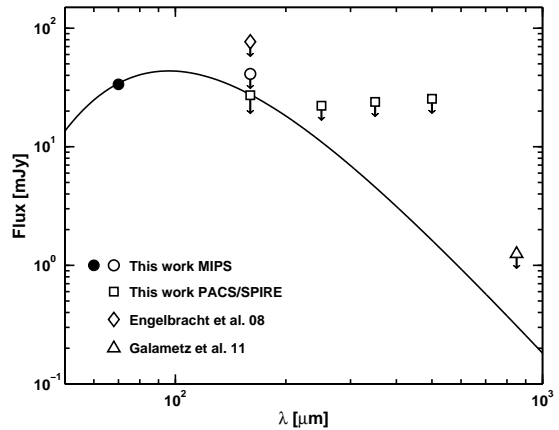}
\caption{A Infrared SED of I~Zw~18. Open symbols represent 3$\sigma$ upper limits. 
The circles show the new MIPS data. 
The squares show the PACS and SPIRE data. 
The open diamond and triangle at 160 and
850~$\mu$m corresponds to upper limits
estimated by \cite{rhc_engelbracht08} and \cite{rhc_galametz11}
respectively.
The solid line corresponds to the modified blackbody fit to the
MIPS 70~$\mu$m detection and the PACS 160~$\mu$m upper limit.
\label{sed}}
\end{figure}

\subsection{Dust Mass}

We use two methods to estimate the dust mass of I~Zw~18. In the first,
we follow the procedure outlined by \cite{rhc_hil83}, assuming an
idealized graybody source with a single temperature.
In the second one we use the \cite{rhc_draine07} (DL07) model. 
The main difference between the DL07 model
and the idealized graybody is that DL07 assumes a grain size distribution 
that reproduces the observed wavelength-dependence extinction in the Milky Way, 
and consequently a distribution of temperatures.
Giving the extreme nature of I~Zw~18, it is not clear that either
model is exactly applicable. Nonetheless, we used them so we
can make a consistent comparison to larger samples of galaxies.

\subsubsection{Modified Blackbody Model}
For an idealized cloud, the dust mass is estimated by fitting its far
infrared spectrum as the product of a blackbody spectrum
($B_{\lambda,T}$) and a mass absorption coefficient
($\kappa_{\lambda}$). The absorption coefficient varies with
wavelength as the negative power of the grain emissivity index
($\kappa_{\lambda} \propto \lambda^{-\beta}$, where $\beta$ represents
the emissivity index). Then, for a cloud that is optically thick to
starlight and optically thin to far infrared emission, the dust mass
M$_{Dust}$ is given by the following expression:

\begin{equation} \label{eq1}
M_{Dust} = \frac{F_{\lambda}D^{2}}{\kappa_{\lambda}B_{\lambda,T}} \ \ ,
\end{equation}

\noindent 
where D is the distance to the galaxy, F$_{\lambda}$ is the observed
flux and $B_{\lambda,T}$ is the blackbody intensity. 
The flux at any point on a graybody spectrum is F$_{\lambda} \propto
B_{\lambda,T} \lambda^{-\beta}$, with $\beta$ independent of
wavelength. Therefore, we can solve for the color temperature
(T$_{70/160}$) using the ratio F$_{70}$ / F$_{160}$.

Measured values for $\kappa_{\lambda}$ at 250 $\mu$m ($\kappa_{250}$)
span the range $\approx 5 - 15$ cm$^{2}$~g$^{-1}$\citep{rhc_alton04}, and
commonly used values for $\beta$ are $1-2$ depending on the
environment. For this work we adopt $\kappa_{250} =$ 9.5
cm$^{2}$~g$^{-1}$ and $\beta = 1.5 $; these are commonly used values 
for low metallicity galaxies\citep[e.g.][]{rhc_leroy07b}.
Using $\beta = 1$ or
2 changes our dust mass limits by $\sim10\%$. 

Fig.~\ref{sed} shows the spectral energy distribution (SED) of I~Zw~18.
The source is only detected at 70~$\mu$m. At longer wavelengths
each point corresponds to a $3\sigma$ flux upper limits. Among these limits,
the 160~$\mu$m upper limit represents the strongest constraint on 
the I~Zw~18 SED. Thus, based on the 70 and
160~$\mu$m emission,  the modified blackbody
spectrum model constrains the dust temperature to be T$_{70/160} >
33.7$~K. This translates into a predicted 850 $\mu$m flux of 0.28~mJy,
compatible with the observed upper limit of 1.25~mJy \citep{rhc_galametz11}.  
Combining our temperature lower limit with
the 70 $\mu$m flux, Eq.~\ref{eq1} yields a dust mass of M$_{Dust} <
3.2~\times~10^{3}$~M$_{\odot}$. If we use the
MIPS 160~$\mu$m upper limit instead of the PACS upper limit,
we measure a temperature limit T$_{70/160} >
29.8$~K and a dust mass a factor of $\sim$2 higher, 
i.e., M$_{Dust} < 6.9~\times~10^{3}$~M$_{\odot}$

\subsubsection{Draine \& Li Model}

For a detailed description we refer to DL07 and
\cite{rhc_draine07b}. Essentially, DL07 models characterize the dust
as a mixture of carbonaceous and amorphous silicate grains with size
distributions chosen to match the observed extinction in the Milky Way.
To characterize the intensity of the radiation that is heating the dust,
the model adopts the spectrum of the local interstellar radiation field 
(this may not be a good approximation for I~Zw~18, a starburst system 
characterized by high intensity radiation fields and low metallicity).
In DL07, most dust is heated by the interstellar radiation
field, and a small fraction is heated by stronger radiation fields
associated with star formation. 

\begin{table*}
\begin{center}
\caption{Derived dust properties based on the DL07 model\label{tbl-1}}
\begin{tabular}{ccccccc}
\\
\tableline\tableline
Object & M$_{Dust}$ (M$_{\odot}$) & U$_{min}$ & $\langle U \rangle$ & $\gamma$ (\%) & $f_{PDR} (\%)$& T$_{Umin}$ (K) \\
\tableline
I Zw 18 & $ < 1.1 \times 10^{4}$ & $ > 8.8 $ & $>$ 21.1 & $>$ 13.1 & $>$ 48.9 & $>$ 24.4 \\
\tableline
Mrk 33 & $2.9 \times 10^{6}$ & 4.0 & 14.3 & 11.8 & 47.6 & 21.4 \\
Tol 89  & $2.8 \times 10^{6}$ & 2.0 & 3.6 & 6.4 & 33.2 & 19.1 \\
NGC 3049 & $5.5 \times 10^{6}$ & 3.0 & 5.1 & 6.1 & 32.6 & 20.4 \\
SINGS\tablenotemark{a} & 1.5 $\times 10^{7}$ & 1.5 & 1.7 & 0.9 & 7.8 & 18.2 \\
\tableline
\end{tabular}
\end{center}
\tablenotemark{a} Median values for 48 SINGS galaxies (Table 5, \citealp{rhc_draine07b})
\end{table*}

We caution the reader that estimating dust masses based on broadband infrared fluxes, as we do here,
is a poorly constrained technique; there are very few data points compared to the number of parameters in the model.
The DL07 model uses five parameters to characterize the emission from dust in galaxies: 
$M_{Dust}$, $U_{min}$, $U_{max}$ $\gamma$,  and $\alpha$. The dust mass is represented by $M_{Dust}$.  
$U_{min}$ represents the interstellar radiation field heating the diffuse ISM, and $U_{max}$ 
represents the upper limit on the interstellar radiation field. The starlight heating the dust is
described using the dimensionless parameter $U$, that by definition is always between 
$U_{min}$ and $U_{max}$. The value $U=1$ is the local interstellar radiation in the Milky Way.  
The parameter $\gamma$ represents the fraction of gas that is exposed to
strong radiation fields with intensities in the range $U_{min} < U < U_{max}$. Finally, 
$\alpha$ characterizes the distribution of starlight intensities. 
In practice we fix two of these parameters, $\alpha$ and $U_{max}$. We adopt the values set by 
\cite{rhc_draine07} of $U_{max} = 10^{6}$ and $\alpha = 2$. Therefore three parameters are free in the model 
($M_{Dust}$, $U_{min}$, and $\gamma$). We remind the reader that we constrain this model with only four broadband fluxes, at
8, 24, 70 and 160~$\mu$m. We can also use the returned values of these parameters to 
calculate a temperature for the majority of the dust grains ($T_{Umin}$), the fraction of dust luminosity that originates 
in photon dominated regions ($f_{PDR}$), and the dust-weighted mean starlight intensity ($\langle U \rangle$). 
 
 

\cite{rhc_mmateos09} derived empirical fits relating a grid of DL07 emission 
model outputs to the {\it Spitzer} fluxes. 
In particular, the DL07 parameters $M_{Dust}$, $\gamma$, $\langle U \rangle$, 
and $f_{PDR}$ can all be derive using measurements at 8, 24, 70 and
160 $\mu$m. We can calculate $U_{min}$ using $\langle U \rangle$ and $\gamma$
according to Eq. (33) in DL07.
For the I Zw 18 fluxes at 8 and 24~$\mu$m we used the
values measured by \cite{rhc_engelbracht08} of 0.47 and 6.28 mJy respectively. 
The agreement between the dust mass derived using the empirical fits and
DL07 models is very good, with a scatter of about 9\% and an offset
of $+5\%$. The DL07 dust masses are strongly dependent on
R$_{70} \equiv \langle \nu F_{\nu} \rangle_{70} / \langle \nu F_{\nu}
\rangle_{160}$, with M$_{Dust}~\propto$ ~R$_{70}^{-1.8}$. R$_{70}$ is
sensitive to the temperature of the largest grains dominating the FIR
emission, and any new constraint or detection at 160~$\mu$m will
strongly affect the resulting dust mass. Essentially, the smaller the 160~$\mu$m flux, 
the hotter the temperature and thus the less dust is needed to produce 
the observed 70~$\mu$m emission.

The derived dust properties for I Zw 18 are summarized in
Table~\ref{tbl-1}. Median values for 48 SINGS galaxies analyzed by
\cite{rhc_draine07b} and three starburst system out of the same sample
are included for comparison. The lower limits obtained for 
$U_{min}$, $\langle U \rangle$, $\gamma$ and $f_{PDR}$ in I~Zw~18 are high
compared to the mean values in the SINGS sample 
\citep{rhc_draine07b}. The high radiation intensity environment of I Zw 18 is
comparable to starbursting systems like Mrk~33, Tol~89 and NGC~3049,
as also found by \citet{rhc_wu07}. We find that the DL07 model yields
a mass upper limit of M$_{Dust} <$~1.1~$\times$~10$^{4}$~M$_{\odot}$.
Just like the modified blackbody case, if we use the
MIPS 160~$\mu$m upper limit instead of the PACS upper limit,
we measure a dust mass a factor of $\sim$2 higher, i.e., 
M$_{Dust} <$~2.6~$\times$~10$^{4}$~M$_{\odot}$.

\subsection{Comparison of Dust Masses}

The DL07 dust mass upper limit is a factor $\sim$3.5 larger than the
dust mass estimated using the modified blackbody model. 
DL07 model treats dust emission as an ensemble of 
dust grains at different temperatures that includes 
larger masses of dust at colder temperatures than 
what is predicted by the single temperature fit.
Therefore, it is not surprising that this model 
generates a higher dust mass than the modified blackbody model.
However, the fact that these two measurements are not extremely different
increases our confidence in the dust mass limit, which we conservatively take to be
that resulting from the DL07 model. 

It is interesting to compare this result to other modeling efforts for
I~Zw~18 and low metallicity galaxies. In particular,
\cite{rhc_galametz11} determine dust masses in a large sample of
galaxies with literature data, using full spectral energy distribution
modelling based on the \citet{rhc_zubko04} grain model.  They find
that the inclusion of submm-wave data tends to drive dusty galaxies
toward lower dust masses, while for low metallicity galaxies the
inclusion of submm-wave constraints yields higher dust mass
predictions than those from far-infrared alone. By contrast
\citet{rhc_draine07b} found their masses to be robust to the inclusion
of submm-wave data.  This is in part driven by modeling choices, in
particular the inclusion of a minimum radiation field or the
interpretation of submm-wave excess \citep{rhc_israel10,rhc_bot10} as
caused by cold dust. The latter appears to be associated to low
metallicities, and although recent studies suggest that it is not
caused by cold dust \citep{rhc_galliano11}, it remains a large
systematic uncertainty in dust mass determinations. Also note that for
dwarf galaxies the SCUBA data included by \citealt{rhc_galametz11} is
much deeper than the data used by
\citealt{rhc_draine07b}. Observational biases and limitations have a
non negligible impact on the interpretation of the observed trends in
dust-to-gas ratio versus metallicity.

\citet{rhc_galametz11} use a 160~$\mu$m flux upper limit of 76.8~mJy 
\citep{rhc_engelbracht08}. For submm-wave data they use a 850~$\mu$m 
flux upper limit of 1.25~mJy \citep{rhc_galliano08}. They measure an
upper limit to the dust mass of $1.1\times10^{5}$~M$_{\odot}$ scaled
to our adopted distance.  Our dust mass upper limit is 
an order of magnitude smaller than that of \cite{rhc_galametz11}.
Using the DL07
model and the data in \cite{rhc_galametz11} we would obtain an upper
limit of $1.0\times10^{5}$~M$_{\odot}$. This is nearly equivalent to
the dust mass \cite{rhc_galametz11} finds, highlighting the fact that
the mass limits are likely reasonably robust to the choice of models. Therefore,
our lower dust mass limit is due to the tighter flux limits at 160 $\mu$m. 

\begin{figure*}
\epsscale{1.0}
\plotone{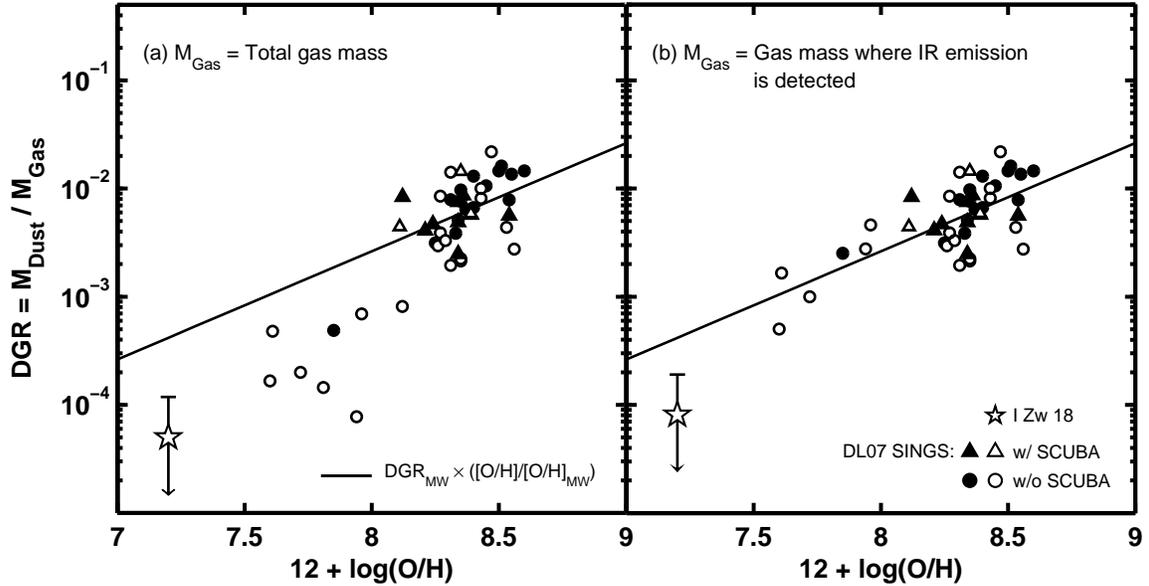}
\caption{ Dust-to-gas mass ratio is shown as a function of oxygen abundance. Open symbols
correspond to upper limits (3$\sigma$). I~Zw~18 DGR is measured
based on the PACS 160~$\mu$m flux upper-limit and is shown as an open star.
The upper error bar corresponds to the value of the DGR if estimated
based on the MIPS 160~$\mu$m flux (3$\sigma$) upper-limit, 
which may be more appropriate if the source is extended at 160 $\mu$m 
over scales larger than 12\arcsec. 
Circles and triangles correspond to SINGS galaxies with and 
without SCUBA fluxes, respectively.
The dust and gas mass values are from Tables 4 and 5 in \cite{rhc_draine07b}. 
The values in the left panel represent the global DGR when estimated using
the total gas mass of the galaxy. The right panel shows the DGR values
when they are estimated ``locally", using only the gas mass in the region where
the IR emission is detected. The solid line shows a linear
scaling between DGR and metallicity normalized to match the Milky Way values (eq. 13, \citealt{rhc_draine07b}). 
\label{dgr}}
\vskip 5 pt
\end{figure*}

\subsection{Gas Mass}

The total \hi\ mass of I~Zw~18 is $M_{HI}=2.3\times10^{8}$~M$_{\odot}$
\citep{rhc_vanzee98}. The molecular content of I~Zw~18, however, remains
unknown since no CO emission has been detected. 
Our new upper limit on the CO
$J=1\rightarrow0$ luminosity of I~Zw~18 is $L_{\rm CO} \leq
10^5$~K~km~s$^{-1}$~pc$^2$ (4$\sigma$), which corresponds to $M_{\rm
H2} \leq 450,000$ M$_\odot$ for a standard conversion factor 
($\alpha_{CO} = 4.5~M_{\odot}$~(K~km~s$^{-1}$~pc$^2$)$^{-1}$). Note
that our luminosity is similar to that quoted by \citet{rhc_leroy07}
because we adopt here a larger distance (for matched smoothing and
assumptions we improve the sensitivity of that study by a factor of
2). Using this Milky Way based conversion factor, M$_{\rm H2}$ is
at most $0.2\%$ of the total gas mass.


There is no reason to expect that the Milky Way based conversion factor between
CO luminosity and H$_{2}$ mass applies to low metallicity galaxies like I~Zw~18.
In the Local Group, $\alpha_{CO}$ is a strong function of metallicity \citep{rhc_leroy11}.
\cite{rhc_genzel12} derive a correlation between oxygen abundance and 
conversion factor, $\alpha_{CO}$. Applying their formula  
to a galaxy with the metallicity of I~Zw~18, we find a conversion factor
$\alpha_{CO} \approx 477.5$~M$_{\odot}$~(K~km~s$^{-1}$~pc$^2$)$^{-1}$. This factor
is $\sim100$ times larger than $\alpha_{CO}$ in a typical spiral galaxy. 
Using this conversion factor we calculate a molecular gas mass upper limit 
of $M_{\rm H2} \leq 4.8\times10^{7}$ M$_\odot$. This $M_{\rm H2}$ is
$\sim20\%$ of the total gas mass.

There is some evidence that, at low metallicities, the star formation
activity may be a better indicator of the molecular mass than the CO
emission \citep{rhc_krumholz11,rhc_bolatto11,rhc_schruba11}.  The H$\alpha$ flux of
I~Zw~18 suggests a recent star formation rate (SFR) of $\sim
0.1$~M$_\odot$~yr$^{-1}$. In large star-forming galaxies, a typical
H$_2$-to-SFR ratio (H$_2$ depletion time) is $\sim 1-2$~Gyr
\citep{rhc_bigiel11}. The H$_2$ mass corresponding to
this amount of star formation in such a galaxy would thus be $\sim
10^8$~M$_\odot$. 
Given the level at which star formation obviously
dominates the morphology and ISM conditions in I~Zw~18 we strongly
suspect that this kind of equilibrium assumption very much
overestimates the H$_2$, which will have been dissociated or otherwise
destroyed by the recent burst. Nonetheless, even in this limit the
H$_2$ only makes up $\sim 30\%$ of the integrated gas mass. The similarity 
with the results obtained from applying the \cite{rhc_genzel12} correlation is 
not surprising, since the underlying assumption is the same. Because of
its uncertainty, we do not include the H$_2$ correction in the following calculations.

\subsection{Dust-to-Gas Mass Ratio and Metallicity}

\cite{rhc_draine07b} finds that the DGR
changes significantly depending on whether the dust mass is compared
to the total gas mass or only the gas mass enclosed in the aperture
where the infrared emission is measured. 
For example, IC~2574 (a dwarf galaxy in the SINGS sample),
 only 19\% of the \hi\  gas mass is enclosed in the area where the infrared 
 emission is detected \citep{rhc_walter07}. 
We take the point of view that these are two valid definitions of
the DGR: global, or local where the dust emission is detected. Using the
total \hi\ mass from \cite{rhc_vanzee98}, we measure an upper limit
for the global DGR~$\lesssim5\times10^{-5}$. Using instead the \hi\ mass
enclosed in the area were we measure the 160~$\mu$m flux upper limit
(62\% of the total \hi\ mass), yields a local DGR~$\lesssim8.1\times10^{-5}$.

Figure~\ref{dgr} shows the DGR as a function of oxygen abundance for
I~Zw~18 and a subsample of SINGS galaxies. Open symbols indicate
3$\sigma$ upper limits.  The solid line represents a linear scaling of
the DGR with metallicity.  This linear relation assumes that the
abundances of all heavy elements are proportional to the oxygen
abundance and that the same fraction of all heavy elements are in
solid form as in the Milky Way \citep{rhc_draine07b}.  The I~Zw~18 DGR
upper limit is primarily driven by the upper limit in the dust mass,
while in the SINGS galaxies the upper limits are due to lower limits
in the gas mass (due to the non inclusion of H$_2$).  As we discussed
in Section 3.1, we obtain different dust masses for I~Zw~18 depending
on the assumption we make about the distribution of the 160~$\mu$m
emission (point-like with PACS and extended with MIPS).  The open star
shows the DGR of I~Zw~18 when we assume point-like emission, while the
upper limit of the bar shows the DGR when we assume extended emission.
For the SINGS galaxies, the dust and gas masses are from
\cite{rhc_draine07b} and the metallicities are from
\cite{rhc_moustakas10}. Note that \citet{rhc_draine07b} computes
molecular gas masses assuming a fixed $X_{CO}$ factor of
4~$\times$~10$^{20}$~cm$^{-2}$ ~(K~km~s$^{-1}$)$^{-1}$ following
\citep{rhc_blitz07}.
We show SINGS galaxies with and without measured SCUBA fluxes as
triangles and circles respectively, as \cite{rhc_draine07b} find that
the dust mass estimates with and without SCUBA data can differ by a
factor of $\sim$2.

The left panel of Figure~\ref{dgr} shows the global DGR, 
estimated using the total gas masses.
The right panel shows how the DGR changes when 
estimating it locally in low metallicity systems,
including I~Zw~18, by using the gas mass enclosed
in the region where the infrared emission is detected.
For SINGS galaxies with metallicities 12~+~log(O/H)~$\gtrsim$~8.1, 
the total DGR seems to agree within a factor of $\sim$2 with
a linear relationship between DGR and metallicity.
Low metallicity galaxies do not seem to follow
the same linear correlation that includes the Milky Way DGR. 
The  I~Zw~18 global DGR falls below the linear scaling by a factor of $\sim8$.
The right panel of Figure~\ref{dgr} shows the local DGR.  The DGRs of
the SINGS low metallicity systems scale up and appear consistent with
the linear relationship within a factor of $\sim$2, although most of
the low metallicity points are only upper limits. For I~Zw~18,
however, the local DGR falls below the linear scaling by a factor of
$\sim5$. Therefore, our dust mass limits for I~Zw~18 suggest a
breakdown of the linear relationship between DGR and metallicity at
very low metallicities.
Note further that, at least in terms of the global DGR, I~Zw~18 seems
to continue the trend found for other low metallicity galaxies.

We show in Fig.~\ref{dgr} that only one of the seven SINGS galaxies with 
12~+~log(O/H)~$\lesssim$~8.1 has a DGR that is not an upper limit. 
It may be possible that the local DGR of this  system is higher than other low metallicity galaxies and the trend is really steeper than linear,
as our result for I~Zw~18 and other studies suggest \citep{rhc_lisenfeld98, rhc_mmateos09}.
Clearly more work is needed to determine, robustly, wether low metallicity galaxies do or do not follow the linear scaling shown in Fig.~\ref{dgr}. 

Note that the abundance of oxygen may not be the 
correct abundance to refer to. Indeed, the abundances of 
refractory elements that constitute the bulk of the dust such as carbon or silicon
are likely more relevant to establishing the DGR.
\cite{rhc_garnett99}, for example,
find a trend of increasing C/O with O/H for a sample of 
irregular and spiral galaxies observed with HST.
This could suggest that the nonlinear trend of DGR
with metallicity is really an artifact of using O/H
as a proxy for metallicity, and the relation could become
more nearly linear when plotted against C/H. \cite{rhc_garnett99} 
find a gas-phase abundance of C in
I~Zw~18 that is significantly higher than that
predicted by the extrapolation of the observed C/O vs. O/H trend in 
low metallicity irregular galaxies. In fact, C/O in
I~Zw~18 is only 0.3~dex lower than Solar. This 
is barely enough to reconcile our limits on the 
local DGR with a linear trend with C/H, and probably
not enough to explain our low global DGR, but it
certainly goes in the right direction.


\section{Conclusions}

In this work we study I~Zw~18 using data from {\it Spitzer}, {\it Herschel Space Telescope} and 
IRAM Plateau de Bure Interferometer. We reduce the flux upper limit at 160~$\mu$m by a factor of $\sim$3 and
the CO $J=1\rightarrow0$ flux upper limit by a factor of $\sim$2 compared to previous measurements. 
Combining these observations with
the dust emission model from \cite{rhc_draine07}, we constrain the
dust mass to be M$_{dust}<1.1\times10^{4}$~$M_{\odot}$. We note that
any dust mass measurement relies on assumptions about the mass
emissivity of dust grains in the interstellar medium, with the
important associated systematic uncertainties.  We find a global
dust-to-gas mass ratio of M$_{dust}/$M$_{gas}<5.0\times10^{-5}$, while
the ratio measured in regions where the 70~$\mu$m emission peaks is
M$_{dust}/$M$_{gas}<8.1\times10^{-5}$.

These measurements are suggestive that low metallicity galaxies do not follow
the same linear relationship between metallicity and DGR as typical local spirals.
At face value our DGR upper limit is inconsistent with the hypothesis
that the fraction of heavy elements incorporated into dust is the same
in high metallicity galaxies (such as the Milky Way) and in extremely
low metallicity galaxies (such as I~Zw~18). There are other scenarios,
however, that can produce a break or non-linear power-law relationship
between DGR and metallicity. For instance, models that include more
detailed physical processes such as the production and destruction of
dust by supernovae, removal of dust through outflows from galaxies,
and dust production in the envelopes of stars
\citep[e.g,][]{rhc_lisenfeld98,rhc_edmunds01,rhc_hirashita02} may
yield non-linear relations. Much more work is needed with sensitive maps
of low metallicity galaxies, like I~Zw~18, to better understand the realtionship
between DGR and metallicity.

\acknowledgments

We thank the referee for his/her comments, 
which helped us to improve the paper considerably.
We thank L. van Zee, and S. Janowiecki for discussions during this
project. This work is based on observations made with the {\em Spitzer
Space Telescope}, operated by JPL/Caltech under a contract with
NASA. Support for this work was provided in part by NASA JPL-1314022,
NSF AST-0838178, NSF AST-0955836, and a Cottrell Scholar award from
the RCSA.  IRAM is supported by INSU/CNRS (France), MPG (Germany) and
IGN (Spain).  The National Radio Astronomy Observatory is a National
Science Foundation facility operated under a cooperative agreement by
Associated Universities, Inc.

{\it Facilities:} \facility{Spitzer}, \facility{Herschel}, \facility{VLA}, \facility{PdBI}.


\end{document}